\documentclass[prb,aps,showpacs,floatfix]{revtex4}
\usepackage{graphicx}
 
\begin{document}

\title{Transverse spin waves in isotropic ferromagnets} 
\author{V. P. Mineev}
\affiliation{Commissariat \'a l'Energie Atomique, DSM/DRFMC/SPSMS
38054 Grenoble, France and Landau Institute for Theoretical Physics,
Chernogolovka, Russia }

\date{July 9, 2005}

\begin{abstract}
The comparison of transverse spin wave spectra and its attenuation in 
Heisenberg ferromagnet
and in ferromagnetic Fermi liquid as well in polarized Fermi liquid is 
undertaken.
The transverse spin waves frequency in polarized paramagnetic Fermi 
liquid as well in a Fermi liquid
with spontaneous magnetization is found to be proportional to $k^2$ with 
complex diffusion
coefficient such that the damping  has a finite value proportional to 
the scattering rate of quasiparticles at $T=0$.

This behavior of
polarized Fermi liquid contrasts with the behavior of Heisenberg
ferromagnet in hydrodynamic regime where the transverse spin wave
attenuation appears in terms proportional $k^4$.

The reactive part of diffusion coefficient in paramagnetic state at $T=0$ 
proves to be inversely
proportional to magnetization whereas in ferromagnetic state it is 
directly proportional to magnetization.
The dissipative part of diffusion coefficient  at $T=0$ in paramagnetic 
state is polarization
independent, whereas in ferromagnetic state it is proportional to square 
of magnetization.  Moreover, the spin wave spectrum in ferromagnetic
Fermi liquid proves to be unstable that demonstrates the difficulty of
the Fermi liquid description of itinerant ferromagnetism.
\end{abstract}

\maketitle
    
\bigskip 

\section{Introduction}

The significant role in physics belong to the simplest models taking in
consideration the essential features of physical phenomena.  The
particular examples of such kind models in the magnetism theory are
the models of isotropic ferromagnets.  There are two type of isotropic
ferromagnets.  First , it is the system of localized moments occupying
the sites of some crystal lattice with Heisenberg exchange interaction
between them
\begin{equation}
\hat H_{Heis}¥=-\sum_{\langle ij\rangle}¥J_{ij}¥{\bf S}_{i}¥{\bf S}_{j}¥.
\label{e1}
\end{equation}
Second, it is Stoner-Hubbard ferromagnet formed of fermi-particles moving
over a crystal lattice with local repulsive interaction between the
particles with opposite spins
\begin{equation}
\hat H_{S-H}¥=\sum_{\langle
ij\rangle,\sigma}¥t_{ij}¥a^{+}¥_{i\sigma}¥a_{j\sigma}¥+
U\sum_{i}¥n_{i\uparrow}¥n_{i\downarrow}¥=\sum_{\langle
ij\rangle,\sigma}¥t_{ij}¥a^{+}¥_{i\sigma}¥a_{j\sigma}¥+\frac{NU}{2}-
\frac{2U}{3}\sum_{i}¥{\bf S}_{i}¥^{2}¥.
\label{e2}
\end{equation}
The latter system can be treated on more phenomenological level as
Landau Fermi liquid with short range interaction between
quasiparticles with opposite spins.  The polarization in the Fermi liquid can
be either spontaneous, this is the case due to Pomeranchuk type
instability at $1+F^{a}¥_{0}¥<0$, or caused by external field or
pumping in a paramagnetic state.  We shall discuss below both of these
possibilities.

It is obvious that the systems described by the Hamiltonians (\ref{e1})
and (\ref{e2}) are isotropic that means the invariance of the
Hamiltonians under simultaneous homogenious rotation of all the spins.
Unlike this common feature we shall see that the systems of localized and
itinerant spins posess quite different properties.  A comparison of
them it is the goal of this paper.  We shall be particularly
interested in the transverse spin dynamics and present here results of
its phenomenological and microscopic theory approach considerations.

\section{Transverse spin waves in Heisenberg ferromagnet}

The treatment of transverse spin waves hydrodynamics has been undertaken by 
several authors (see eg \cite{1,2,3}).  We rederive it here in
somewhat different manner.  With this purpose instead of Hamiltonian
(\ref{e1}) we shall use a phenomenological expression for the free
energy
\begin{equation}
F=F({\bf M})-{\bf M}{\bf H_{0}¥}+ \frac{a}{2}(\nabla_{i}¥M_{\alpha}¥)^{2}¥
\label{e3}
\end{equation}
where the minimum of $F({\bf M})$ gives the equlibrium value of magnetization 
 ${\bf M}$ in ferromagnetic state and ${\bf H_{0}¥}$ is an external
 magnetic field.  Being interested in the dispersion law of small transverse
 vibrations of magnetzation $\delta M_{\alpha}¥=
 e_{\alpha\beta\gamma}¥\Theta_{\beta}¥M_{\gamma}¥$, where $\Theta$ is a
 vector of infinitizemal rotation lying in the plane perpendicular to
 ${\bf M}$ we rewrite the free energy in terms of these angles
\begin{equation} 
 F=F({\bf M})-{\bf M}{\bf H_{0}¥}+ \frac{a}{2}{\bf
 M}^{2}¥(\nabla_{i}¥\Theta_{\alpha}¥)^{2}¥.
\label{e4}
\end{equation}
Then, by introducing the magnetization current as
\begin{equation} 
J_{i\alpha}¥=-\frac{\delta F}{\delta \nabla_{i}¥\Theta_{\alpha}¥}=
-a({\bf M}\times\nabla_{i}¥{\bf M})_{\alpha}¥,
\label{e5}
\end{equation}
we obtain the equation of motion of magnetization or the equation of
spin density conservation
\begin{equation} 
\frac{\partial {\bf M}}{\partial t}+\frac{\partial {\bf J}_{i}¥}
{\partial x_{i}¥}-{\bf M}\times\gamma{\bf H_{0}¥}=0
\label{e6}
\end{equation}
known as Landau-Lifshits equation \cite{4}.
The simple derivation from eqns (\ref{e5}), (\ref{e6}) results in dispersion
law of linear transverse spin waves
\begin{equation} 
\omega=\omega_{L}¥+aMk^{2}¥.
\label{e7}
\end{equation}
Here $\omega_{L}¥=\gamma H_{0}¥$ is the Larmor frequency.

So, the reactive part of spin waves dispersion proves to be directly
proportional to the magnetization value.  This general property of
Landau-Lifshits equation is sometimes formulated as resulting of
finite domain wall rigidity.

The dissipation can be also taken into
consideration.  It is only necessary to generalize the spin current
expression
\begin{equation} 
J_{i\alpha}¥=-\frac{\delta F}{\delta \nabla_{i}¥\Theta_{\alpha}¥}+
J_{i\alpha}¥^{diss}¥,~~~~~J_{i\alpha}¥^{diss}¥=
-\frac{\delta R}{\delta \nabla_{i}¥\Theta_{\alpha}¥},
\label{e8}
\end{equation}
where 
\begin{equation} 
R=b
e_{\alpha\beta\gamma}¥M_{\alpha}¥\nabla_{i}¥\nabla_{j}¥\Theta_{\beta}¥
\nabla_{i}¥\nabla_{j}¥\Theta_{\gamma}¥
\label{e9}
\end{equation}
is the dissipation function which according to general rules is chosen being
quadratic on gradients of spin velocity $\nabla_{i}¥\Theta_{\alpha}¥$ 
(it is variable conjugated to the spin current) and such that
dissipative part of the spin current is an odd function in respect of time
inversion
\begin{equation} 
J_{i\alpha}¥^{diss}¥(-t)=-J_{i\alpha}¥^{diss}¥(t).
\label{e10}
\end{equation}
Taking into account the dissipative spin current we obtain from
Landau-Lifshits equation the transverse spin waves dispersion law with
dissipation
\begin{equation} 
\omega=\omega_{L}¥+aMk^{2}¥-ib k^{4}¥.
\label{e11}
\end{equation}
The microscopic calculation \cite{5,6}  gives the value of coefficient
$b\propto (\ln T/k^{2}¥)/|M|^{3}¥$ meaning the nonanalytic wave vector
dependence of dispersion law.

In conclusion of this Section we stress that all the results found
here are valid in hydrodynamic or local equilibrium regime that is
under the following condition
\begin{equation} 
aMk^{2}¥\tau \ll 1
\label{e12}
\end{equation}

\section{Transverse spin waves in polarized Fermi liquid }

In spin polarized Fermi liquid the equation of spin density conservation
is still valid
\begin{equation} 
\frac{\partial {\bf M}}{\partial t}+\frac{\partial {\bf J}_{i}¥}
{\partial x_{i}¥}-{\bf M}\times\gamma{\bf H_{0}¥}=0
\label{e13}
\end{equation}
but the spin current density has the following form
\begin{equation} 
{\bf J}_{i}¥=-D^{\prime}¥\nabla_{i}¥{\bf M}+D^{\prime\prime}¥~\hat {\bf
m}\times\nabla_{i}¥{\bf M},
\label{e14}
\end{equation}
where $\hat {\bf m}={\bf M}/M$. Here the second term has the same
structure as the reactive part of spin current in Heisenberg
ferromagnet (\ref{e5}).  It is time reversal invariant while the first
term describes the dissipative and odd in respect of time reversal
current.  Such a term is absent in Heisenberg ferromagnet but it is
always present in spin polarized Fermi liquid even at absolute zero
temperature as diffusion current in the solution of two liquids with
up and down spins.

For the weakly polarized paramagnetic Fermi liquid the equations
(\ref{e13})-(\ref{e14}) has been derived from semiclassical Landau-Silin
kinetic equation \cite{7} by A.Leggett \cite{8}.  Then the exact form of
reactive and dissipative part of diffusion constant has been found in
frame of the same approach \cite{9} with general form of two particle collision 
integral in weakly polarized Fermi liquid with arbitrary relationship
between temperature $T$ and polarization $\gamma H$ .
Finally the expressions applicable both for the description of spin
dynamics in paramagnetic Fermi liquid with finite polarization and in a 
ferromagnetic Fermi liquid with spontaneous polarzation was found \cite{10}.
 These results are confirmed by derivation of transverse spin wave
 dispersion law in frame of field theoretical methods from the integral
 equation for the vortex function \cite{10}.  It is shown that similar derivation
 taking into consideration the divergency of static transverse
 susceptibility also leads to the same attenuating spin wave spectrum.

The  dispersion law of the transversal spin waves following from
equations (\ref{e13}), (\ref{e14}) is 
\begin{equation}
\omega=\omega_{L}¥+ (D^{\prime\prime}¥-iD^{\prime}¥)k^{2}¥,
\label{e15}
\end{equation}
where $\omega_{L}¥=\gamma H_{0}¥$ is the Larmor frequency,
\begin{equation} 
D^{\prime}¥=\frac{w^{2}¥\tau} {3(1+(C \tau)^{2}¥
)}\cong\frac{w^{2}¥}{3{C}^{2}¥ \tau}
\label{e39}
\end{equation}
is the dissipative part of diffusion coefficient and 
\begin{equation} 
D^{\prime\prime}¥= C\tau D^{\prime}¥\cong\frac{w^{2}¥}{3 C}
\label{e40}
\end{equation}
is its reactive part.  Here the second approximative values of $D^{\prime}¥$
and $D^{\prime\prime}¥$ correspond to the limit $C\tau\gg1$.

The parameter $C$ expresses through the Fermi liquid constants and
integal of the shifted in respect each other spin-up and spin-down
distributions
\begin{equation}
{\bf C}=\frac{\hat {\bf m}}{N_{0}¥}(F_{0}¥^{a}¥ -\frac{F_{1}¥^{a}¥}{3})\int
d\tau \Delta n_{0}¥(\varepsilon) .
\label{e35}
\end{equation}
The value of relative shift $\gamma H$ in paramagnetic Fermi liquid is
determined by the external field, Landau molecular field and in
general nonequlibrium case it is also can be created by pumping.  In the
ferromagnetic Fermi liquid the relative shift exist even in the
absence of external field and it is determined by Fermi liquid (exchange)
interaction. 
The current relaxation time is \cite{9}
\begin{equation}
\frac{1}{\tau}=\frac{m^{*}¥^{3}¥}{6(2\pi)^{5}¥} (2\overline {W_{1}¥}+\overline{W_{2}¥})
\left[(2\pi T)^{2}¥+(\gamma H)^{2}¥\right].
\label{e37}
\end{equation}

For
a weakly polarized fluid $C=(F_{0}¥^{a}¥-F_{1}¥^{a}¥/3)\gamma H$. 
The expression for $w^{2}¥$ depends of state of liquid.  One can find
it analytically in the case of weak polarization \cite{10}.  In a paramagnetic
Fermi liquid it is
\begin{equation} 
w^{2}¥=v_{F}¥^{2}¥(1+F_{0}¥^{a}¥)(1+\frac{F_{1}¥^{a}¥}{3})
\label{e40a}
\end{equation}
where $v_{F}$ is the Fermi velocity in unpolarized liquid.  In a
ferromagnetic Fermi liquid (if an external field is smaller than
spontaneous) 
it is
\begin{equation} 
w^{2}¥=-v_{F}¥^{2}¥(1+\frac{F_{1}¥^{a}¥}{3})
\left(\frac{\gamma H}{4\mu}\right)^{2}¥.
\label{e40b}
\end{equation}

Thus, the reactive part of diffusion coefficient in paramagnetic state at
$T=0$ proves to be inversely proportional to magnetization
\begin{equation} 
D^{\prime\prime}¥=\frac{v_{F}¥^{2}¥(1+F_{0}¥^{a}¥)(1+F_{1}¥^{a}¥/3)}
{3 (F_{0}¥^{a}¥-F_{1}¥^{a}¥/3)\gamma H}
\label{e40c}
\end{equation}
 whereas in ferromagnetic state it is directly proportional to
magnetization
\begin{equation} 
D^{\prime\prime}¥=
\frac {v_{F}¥^{2}¥\gamma H}{3(4\varepsilon_{F}¥)^{2}¥}.
\label{e40d}
\end{equation}
The
latter is in exact correspondence with known result obtained in frame of
Stoner-Hubbard model \cite{11}.

The dissipative part of diffusion
coefficient given by eqn (\ref{e39}) at $T=0$ in paramagnetic state is polarization
independent, whereas in ferromagnetic state it is proportional to the 
square of magnetization.  More important, however, that imagenary part of
dispersion law in Fermi liquid with spontaneous magnetization proves to
be positive.  This means the instability of Fermi liquid with
spontaneous magnetization.  This conclusion is obtained in frame of
linear approximation, another words, for the infinitezimally small
transversal deviations of magnetization.  Does the instability
disappear in nonlinear theory or it is principal lack of itinerant
ferromagnetism description in frame of polarized Fermi liquid theory ? 
This is an open problem.

So, the transverse spin waves frequency in polarized paramagnetic
Fermi liquid as well in a Fermi liquid with spontaneous magnetization is
found to be proportional to $k^{2}¥$ with complex diffusion
coefficient such that the damping at $C\tau\gg 1$ has a finite value
proportional to the scattering rate of quasiparticles at $T=0$.  As it
was pointed out in \cite{9} the latter is in formal analogy with
ultrasound attenuation in collisionless regime.  It is worth noting,
however, that in neglect of processes of longitudinal relaxation the
parameter $\gamma H \tau$ has no relation to the local equilibrium
establishment.

The results (\ref{e15})-(\ref{e40d}) are valid both in hydrodynamic
$Dk^{2}¥\tau \ll 1$ and in collisionless regime $ Dk^{2}¥\tau\gg 1$ so
long
\begin{equation} 
Dk^{2}¥\ll \gamma H
\label{e40e}
\end{equation}
that is the condition
of two moment approximation  for the solution of the kinetic
equation \cite{8}.  This behavior of polarized Fermi liquid contrasts
with the behavior of Heisenberg ferromagnet in hydrodynamic regime
where the transverse spin wave attenuation appears in terms
proportional $k^{4}¥$.

\section{Conclusion}

We discussed the transverse spin waves dispersion in two types of isotropic 
ferromagnetic systems: Heisenberg localized ferromagnet and itinerant polarized
Fermi liquid.  In contrast with the Heisenberg ferromagnet, where spin
wave attenuation appears in terms proportional to the wave vector in
the fourth power, the spin polarized Fermi liquid has attenuation already 
in quadratic in the wave vector terms.  At the phenomenological level
this difference originates from the diffusive current which exists in the
mixture of two spin-up and spin-down Fermi liquids even at zero
temperature.  

Unlike paramagnetic polarized Fermi liquid the spectrum of
transverse spin waves in ferromagnetic Fermi liquid demonstrates the
inherent instability pointing out on troubles of pure Fermi liquid
description of itinerant ferromagnetism.

\end{document}